\documentclass[prl,reprint, amsmath,amssymb]{revtex4-1}
\usepackage{graphicx}
\usepackage{dcolumn}
\usepackage{bm}
\usepackage{mathrsfs}
\usepackage{hyperref}
\usepackage{epstopdf}
\usepackage{amsmath}
\begin{document}
\title{Thermodynamics and elasticity of emergent crystals}
\author{Yangfan Hu}
 \email[Corresponding author.]{huyf3@mail.sysu.edu.cn}
\author{Xuejin Wan}
\affiliation{Sino-French Institute of Nuclear Engineering and Technology, Sun Yat-sen University, 519082, Zhuhai, China}

\begin{abstract}
    Periodic field patterns of atoms and their charges/spins/orbits emerge in crystals, forming novel states of matter called emergent crystals (ECs). In recent years, they are observed in diverse systems such as skyrmion crystals in helimagnets, and periodic ripples in 2D materials. ECs essentially changes the properties of material underneath, and are deformable when subject to various effective fields. A major challenge in application is first to predict what kind of EC will appear in the system of interest, and how to quantify its ``elasticity'' when subject to an effective field. Here we establish the theoretical framework of thermodynamics for deformable ECs, and derive from it the linear constitutive equations when subject to the primary external field. We provide a systematic study on the ECs that may appear in helimagnets induced by the  Dzyaloshinskii-Moriya interaction, and analyze their elasticity when subject to bias magnetic fields. We construct in this work the basis of emergent elasticity, a new branch studying deformable emergent crystals under effective fields.
\end{abstract}
\maketitle

\section{Introduction}
Emergent crystals (ECs) are spatially periodic field patterns emerging from atomic crystals. These field patterns appear due to collective behaviors of atoms or their charges, spins, or orbits, etc. They appear in different material systems, including but not limited to skyrmion crystals (SkX)  \cite{muhlbauer2009skyrmion,bauerle1996laboratory,al2001skyrmions,rossler2006spontaneous,fu2016persistent,nych2017spontaneous,das2019observation} in many different systems, and periodic ripples in 2D materials\cite{bao2009controlled,miro2013spontaneous,kou2015anisotropic}. Specifically, in bulk helimagnets, the noncollinear Dzyaloshinskii-Moriya interaction (DMI)\cite{dzialoshinskii1957thermodynamic,moriya1960anisotropic} permits appearance of various kinds of ECs, including but not limited to Bloch-type SkX \cite{muhlbauer2009skyrmion,yu2010real,yu2011near,seki2012observation}, N\'eel-type SkX \cite{kezsmarki2015neel,kurumaji2017neel,padmanabhan2019optically,bordacs2017equilibrium}, and anti-skyrmion crystals \cite{nayak2017magnetic,koshibae2016theory}. The large variety of ECs that can appear in magnetic materials derives from the anisotropy of DMI  \cite{hoffmann2017antiskyrmions,huang2017stabilization}, which is determined by symmetry of the material. Although existing in distinct systems, ECs generally possess the following features: \textbf{a)} their presence essentially changes various kinds of properties of the underlying material (e.g., SkX changes fundamentally the electronic and magnetic properties of the underlying magnets \cite{ritz2013formation,schulz2012emergent}, while ripples strongly influence the electronic, mechanical and optical properties of the underlying 2D materials \cite{kou2015anisotropic,deng2016wrinkled,quereda2016strong,de2008periodically}). \textbf{b)} their lattice constant and localized field pattern inside the lattices are both sensitive to variation of effective external fields,  usually with an elasticity much softer than that of the underlying material\cite{bao2009controlled,white2014electric,shibata2015large,okamura2016transition}. Combining the two features above, we expect to control the periodicity and field patterns of ECs by varying effective external fields, which in turn controls the properties of the underlying materials. As a result, the existence, stability and elasticity of ECs are the cornerstone to realize novel properties that are tunable by external fields. The key, as it always is for all emergent phenomena\cite{anderson1972more}, is first to determine what kind of ECs will appear in the system of interest, and then to explain its stability and elasticity in terms of properties of the underlying materials and their composing atoms.

Spatial patterns formation has long been the subject of interest in the study of liquid crystal\cite{de1974physics}, where the direction angle of molecules are regarded as a deformable periodic function in space, and in the study of self-organized phenomena in fluids and biological tissues\cite{haken2013synergetics}, where the critical condition for the appearance of spatial patterns due to condensation of soft-modes is analyzed. The thermodynamics of ECs and the corresponding theory of emergent elasticity (i.e., elasticity of ECs) can be established upon the combination of some basic ideas of the physics of liquid crystals\cite{de1974physics} and synergetic\cite{haken2013synergetics}. The state of the system is described by an order parameter vector characterizing the physical quantities that dominate the presence of ECs. In the case of ECs in magnets this vector corresponds to the magnetization vector, and in the case of ripples in 2D materials, this vector corresponds to the displacement vector. Here we focus on the cases where the period of EC considered is a large quantity compared with that of the underlying atomic lattices, for which a continuous field description is appropriate. The appearance of a deformable EC in the system considered means that the order parameter vector should be described by several coupling soft-modes whose wave vectors are variables by external fields. The deformation of ECs is thus explicitly determined by the Fourier magnitudes and the wave vectors of the soft mode. By taking the soft-mode-description of the order parameter vector, we can establish the theory of thermodynamics upon that of the underlying material, and further discuss the emergent elasticity of ECs under effective fields.

In this work, we establish the general framework of thermodynamics and emergent elasticity for deformable ECs when the system is subject to the primary external field (work conjugate of the order parameter vector). Based upon this general framework, we systematically study all ECs that can appear in bulk helimagnets due to presence of the DMI, where a symmetry analysis of the form of DMI and related EC for different point groups is provided. Generally, we find four types of ECs (Bloch SkX, N\'eel SkX, Anti-I SkX, Anti-II SkX), and four mixed states of them when anisotropy of DMI is presented. We then study the emergent elasticity of these ECs under bias magnetic fields, where the elasticity of the 8 types of ECs can be effectively described by two models. We establish in this work the foundation to study the deformation, stability, phase transitions, and elementary excitations \cite{hu2019lagrangian} in various kinds of ECs that may appear in magnetic materials.

The contents of this paper are organized as follow. In section \uppercase\expandafter{\romannumeral2}, we first introduce the emergent strain tensor to describe the deformation of ECs, and then establish the general method to derive the thermodynamic potential of deformable ECs from that of the underlying material. In section \uppercase\expandafter{\romannumeral3}, based on the thermodynamic potential obtained, we derive the linear constitutive equations for ECs when the system is subject to the primary external field of the EC considered. In section \uppercase\expandafter{\romannumeral4}, we systematically study the thermodynamic potential for various kinds of ECs induced by the DMI in magnetic materials, and solve from it the equilibrium field configurations of ECs and their variation with certain thermodynamic parameters. To achieve this, we perform a group theoretical analysis on the functional form of DMI permitted by different crystalline groups, which determines the ECs that may appear. We obtain from our study the Bloch SkX, N\'eel SkX, Anti-I SkX, Anti-II SkX and their mixed states. In section \uppercase\expandafter{\romannumeral5}, we study the emergent elastic property of all kinds of ECs that appear in previous section when subject to bias magnetic fields.

\section{Thermodynamics of deformable emergent crystals}
Consider an EC emerging in an ordinary crystal, the existence of the EC can be described by an order parameter vector field $\mathbf{v}$ of the underlying material. For SkX in helimagnets, $\mathbf{v}$ refers to the magnetization vector; for SkX in ferroelectrics\cite{das2019observation}, $\mathbf{v}$ refers to the polarization vector; and for periodically rippled graphene, $\mathbf{v}$ refers to the displacement vector. The long-range order of the emergent crystalline state requires that $\mathbf{v}$ be expressed by a Fourier series\cite{hu2018wave} (or a group of coupling soft-modes)
\begin{equation}
    \mathbf{v}=\sum_{\mathbf{n}}{{{\mathbf{v}}_{{{\mathbf{q}}_{\mathbf{n}}}}}{{\mathrm e}^{\text{i}{{\mathbf{q}}_{\mathbf{n}}}\cdot \mathbf{a}}}},
\label{1}
\end{equation}
where ${{\mathbf{q}}_{\mathbf{n}}}$ denotes the reciprocal lattice vectors of the emergent crystal. For a $d$-dimensional emergent crystal ($d=1, 2, 3$), ${{\mathbf{q}}_{\mathbf{n}}}={{n}_{1}}{{\mathbf{q}}_{1}}+{{n}_{2}}{{\mathbf{q}}_{2}}+...+{{n}_{d}}{{\mathbf{q}}_{d}},$ where $\mathbf{n}={{\left[ {{n}_{1}},\ {{n}_{\text{2}}},\cdots,{{n}_{d}} \right]}^{\mathrm T}}$ is a vector of integers, and ${{\mathbf{q}}_{1}}$, ${{\mathbf{q}}_{\text{2}}}$, $\cdots$, ${{\mathbf{q}}_{d}}$ are the basic reciprocal vectors.
When the work conjugate of $\mathbf{v}$, the primary external field $\mathbf{X}$, changes, the EC is anticipated to deform. For ECs in magnetic materials, $\mathbf{X}$ refers to the magnetic field; for ECs in ferroelectrics, $\mathbf{X}$ refers to the electric field; and for periodically rippled 2D materials, $\mathbf{X}$ refers to the mechanical forces applied perpendicular to the 2D plane. We assume that due to the emergent deformation, the original coordinates $\mathbf{a}$ map to $\mathbf{r}$, which gives $\mathbf{r}=\mathbf{a}+{{\mathbf{u}}^{e}}$. Here ${{\mathbf{u}}^{e}}$ denotes the emergent displacement vector. Similar to atomic crystals, rigid translation of emergent crystals does not induce any change of energy and is not considered here. To describe a deformable EC, we have to transform $\mathbf{a}$ to $\mathbf{r}$ in Eq. (\ref{1}). According to the theory of solid mechanics \cite{fung2017classical}, there are two possible choices of coordinates. For homogeneous deformation of EC, we have $\mathbf{a}=\mathbf{r}-{{\mathbf{u}}^{e}}(\mathbf{r})=\left[ \mathbf{I}-{{\mathbf{F}}^{e}}(\mathbf{r}) \right]\mathbf{r}$ in the Eulerian coordinates, where ${{\mathbf{F}}^{e}}(\mathbf{r})$ is a matrix with components $F_{ij}^{e}(\mathbf{r})=\varepsilon _{ij}^{e}+\omega _{ij}^{e}$, and $\varepsilon _{ij}^{e}=\frac{1}{2}(\frac{\partial u_{i}^{e}}{\partial {{r}_{j}}}+\frac{\partial u_{j}^{e}}{\partial {{r}_{i}}})$, $\omega _{ij}^{e}=\frac{1}{2}(\frac{\partial u_{i}^{e}}{\partial {{r}_{j}}}-\frac{\partial u_{j}^{e}}{\partial {{r}_{i}}})$ are called the emergent Cauchy's strain tensor and the emergent Cauchy's rotation tensor, respectively. In the Lagrangian coordinates, we have $\mathbf{r}=\mathbf{a}+{{\mathbf{u}}^{e}}(\mathbf{a})=\left[ \mathbf{I}+{{\mathbf{F}}^{e}}(\mathbf{a}) \right]\mathbf{a}$, which gives $\mathbf{a}={{\left[ \mathbf{I}+{{\mathbf{F}}^{e}}(\mathbf{a}) \right]}^{-1}}\mathbf{r}$, where the components of ${{\mathbf{F}}^{e}}(\mathbf{a})$ read $F_{ij}^{e}(\mathbf{a})=E _{ij}^{e}+W _{ij}^{e}$, and $E _{ij}^{e}=\frac{1}{2}(\frac{\partial u_{i}^{e}}{\partial {{a}_{j}}}+\frac{\partial u_{j}^{e}}{\partial {{a}_{i}}})$, $W _{ij}^{e}=\frac{1}{2}(\frac{\partial u_{i}^{e}}{\partial {{a}_{j}}}-\frac{\partial u_{j}^{e}}{\partial {{a}_{i}}})$ are called the emergent Green's strain tensor and the emergent Green's rotation tensor, respectively. As a result, Eq. (\ref{1}) becomes
\begin{equation}
    \mathbf{v}=\sum\limits_{\mathbf{n}}{{{\mathbf{v}}_{{{\mathbf{q}}_{\mathbf{n}}}}}{{\mathrm e}^{\text{i}{{\mathbf{q}}_{\mathbf{n}}}\cdot \left[ (\mathbf{I}-{{\mathbf{F}}^{e}}(\mathbf{r}) \mathbf{r} )\right]}}}
    \label{2}
\end{equation}
in the Eulerian coordinates, and
\begin{equation}
\begin{aligned}
\mathbf{v}=\sum\limits_{\mathbf{n}}{{{\mathbf{v}}_{{{\mathbf{q}}_{\mathbf{n}}}}}{{\mathrm e}^{\text{i}{{\mathbf{q}}_{\mathbf{n}}}\cdot \left[{{ (\mathbf{I}+{{\mathbf{F}}^{e}}(\mathbf{a}) )}^{-1}}\mathbf{r}\right]}}}
\label{3}
\end{aligned}
\end{equation}
in the Lagrangian coordinates. One should notice that in Eqs. (\ref{2}, \ref{3}), the value of emergent elastic strains and emergent rotational angles depend on the choice of wave vectors ${{\mathbf{q}}_{\mathbf{n}}}$. In our formulation, ${{\mathbf{q}}_{\mathbf{n}}}$ are referred to as the undeformed wave vectors, which are determined from the undeformed structure of the EC considered. And the deformed wave vectors are $\mathbf{q}_{\mathbf{n}}^{E}=\left[ \mathbf{I}-{{\mathbf{F}}^{e}}(\mathbf{r}) \right]^\mathrm T{{\mathbf{q}}_{\mathbf{n}}}$ in the Eulerian coordinates and $\mathbf{q}_{\mathbf{n}}^{e}={\left[{\left[ \mathbf{I}+{{\mathbf{F}}^{e}}(\mathbf{a}) \right]}^{-1}\right]^\mathrm T}{{\mathbf{q}}_{\mathbf{n}}}$ in the Lagrangian coordinates. From eqs. (\ref{2}, \ref{3}), in the Eulerian (Lagrangian) coordinates the free energy density of the EC generally takes the form $\phi ({{\pmb{\varepsilon}}^{ea}},{{\mathbf{v}}^{q}},{{\mathbf{X}}^{q}},T)$ ($\phi ({{\mathbf{E}}^{ea}},{{\mathbf{v}}^{q}},{{\mathbf{X}}^{q}},T)$), where for 3D ECs
\begin{equation}
    \begin{aligned}
    {{\pmb{\varepsilon}}^{ea}}={{\left[ \varepsilon _{11}^{e},\ \varepsilon _{22}^{e},\ \varepsilon _{33}^{e},\ \varepsilon _{23}^{e},\ \varepsilon _{13}^{e},\ \varepsilon _{12}^{e},\ \omega _{23}^{e},\ \omega _{13}^{e},\ \omega _{12}^{e} \right]}^{\rm T}},
    \label{4}
    \end{aligned}
\end{equation}

\begin{equation}
\begin{aligned}
{{\mathbf{E}}^{ea}}={{\left[ E_{11}^{e},\ E_{22}^{e},\ E_{33}^{e},\ E_{23}^{e},\ E_{13}^{e},\ E_{12}^{e},\ W_{23}^{e},\ W_{13}^{e},\ W_{12}^{e} \right]}^{\rm T}}.
\label{5}
\end{aligned}
\end{equation}

\begin{figure*}
    \centering
    \includegraphics[scale=0.35]{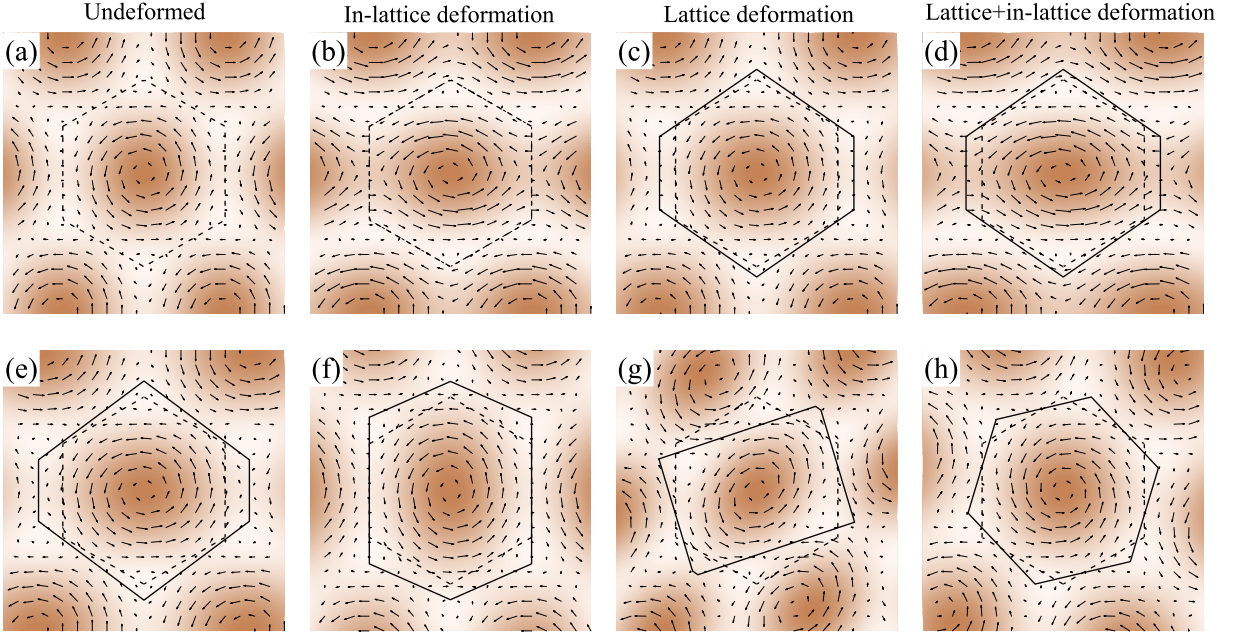}
    \caption{Field patterns of deformed SkX. (a) SkX without deformation. (b) SkX with in-lattice deformation only. (c) SkX with lattice deformation only. (d) SkX with both lattice deformation and in-lattice deformation. (e-h) present four basic modes of lattice deformation of SkX: (e) $\varepsilon_{11}^e=0.3$, (f) $\varepsilon_{22}^e=0.3$, (g) $\varepsilon_{12}^e=0.3$ and (h) $\omega^e=0.3$. The vectors illustrate the distribution of the in-plane magnetization components with length proportional to their magnitude, while the colored density plot illustrates the distribution of the out-of plane magnetization component. The black dashed line plots the undeformed  Wigner-Seitz cell, while the black solid line  plots the deformed cell}
    \label{f11}
\end{figure*}

${{\mathbf{v}}^{q}}$ contains all components of the vectors ${{\mathbf{v}}_{{{\mathbf{q}}_{\mathbf{n}}}}}$ for all possible choices of $\mathbf{n}$, ${{\mathbf{X}}^{q}}$ contains all components of the vectors ${{\mathbf{X}}_{{{\mathbf{q}}_{\mathbf{n}}}}}$ defined by $\mathbf{X}=\sum\limits_{\mathbf{n}}{{{\mathbf{X}}_{{{\mathbf{q}}_{\mathbf{n}}}}}{{e}^{\text{i}{{\mathbf{q}}_{\mathbf{n}}}\cdot \mathbf{r}}}}$, and $T$ denotes the temperature. A fundemental difference between ECs and ordinary crystals is that ECs are composed of localized field patterns instead of point masses. This differece renders two types of deformation that are permitted by ECs: lattice deformation, described by variation of $\pmb \varepsilon^{ea}$, and in-lattice deformation, described by variation of  $\mathbf{v}^q$. For SkX in helimagnets, the difference of these two types of deformation is illustrated in FIG. \ref{f11}. In short, lattice deformation causes simultaneously deformation of the field pattern inside the lattice, while in-lattice deformation does not induce variation of the lattice. One should notice that the Eulerian coordinates is used for the first time in the study of spin waves in SkX\cite{zang2011dynamics}, and will also be used in the following sections. At given temperature $T$ and external field ${{\mathbf{X}}^{q}}$, the field configuration of the EC is obtained by solving the minimization problem of the averaged free energy density $\bar{\phi }=\frac{1}{V}\int_{V}{\phi \left( {{\pmb{\varepsilon }}^{ea}},{{\mathbf{v}}^{q}},{{\mathbf{X}}^{q}},T \right)dV}$.

\section{Linear constitutive equations for emergent crystals}
Now consider an isothermal disturbance of the equilibrium state at given temperature $T$ and external field ${{\mathbf{X}}^{q}}$, where the disturbance is small enough so that it does not lead to any phase transitions and the deviation from the equilibrium state can thus be described by small quantities. To study this deviation, we expand $\bar{\phi }$ in terms of all the independent variables to quadratic terms:

\begin{equation}
\begin{aligned}
\bar{\phi }=& {{\bar{\phi }}_{0}}+\frac{1}{2}{{\left( d{{\pmb{\varepsilon }}^{ea}} \right)}^{\mathrm T}}{{\mathbf{C}}^{e}}d{{\pmb{\varepsilon }}^{ea}}+\frac{1}{2}{{\left( d{{\mathbf{v}}^{q}} \right)}^{\mathrm T}}{{\pmb{\mu}}^{q}}d{{\mathbf{v}}^{q}}\\
& +{{\left( d{{\pmb{\varepsilon }}^{ea}} \right)}^{\mathrm T}}{{\mathbf{g}}^{eq}}d{{\mathbf{v}}^{q}},
\label{6}
\end{aligned}
\end{equation}
where ${{\bar{\phi }}_{0}}$ denotes the undisturbed averaged free energy density, terms with a prefix $d$ denote a small disturbance, and
\begin{equation}
\begin{aligned}
& C_{ij}^{e}={{\left( \frac{{{\partial }^{2}}\bar{\phi }}{\partial \varepsilon _{i}^{ea}\partial \varepsilon _{j}^{ea}} \right)}_{0}},\ \ \mu _{ij}^{q}={{\left( \frac{{{\partial }^{2}}\bar{\phi }}{\partial v_{i}^{q}\partial v_{j}^{q}} \right)}_{0}},\\
 & g_{ij}^{eq}={{\left( \frac{{{\partial }^{2}}\bar{\phi }}{\partial \varepsilon _{i}^{ea}\partial v_{j}^{q}} \right)}_{0}}.
\label{7}
\end{aligned}
\end{equation}
In Eq. (\ref{7}), terms with a subscript 0 take values at the equilibrium state. The linear constitutive equations are derived as:
\begin{equation}
\begin{aligned}
  & d{{\pmb{\sigma }}^{ea}}={{\mathbf{C}}^{e}}d{{\pmb{\varepsilon }}^{ea}}+{{\mathbf{g}}^{eq}}d{{\mathbf{v}}^{q}}, \\
 & d{{\mathbf{X}}^{q}}={{\pmb{\mu }}^{q}}d{{\mathbf{v}}^{q}}+{{\left( {{\mathbf{g}}^{eq}} \right)}^{\mathrm T}}d{{\pmb{\varepsilon }}^{ea}}, \\
\label{8}
\end{aligned}
\end{equation}
where for 3D EC
\begin{equation}
\begin{aligned}
{{\pmb{\sigma }}^{ea}}={{\left[ \sigma _{11}^{e},\ \sigma _{22}^{e},\ \sigma _{33}^{e},\ \sigma _{23}^{e},\ \sigma _{13}^{e},\ \sigma _{12}^{e},\ \Gamma _{23}^{e},\ \Gamma _{13}^{e},\ \Gamma _{12}^{e} \right]}^{\mathrm T}},
\label{9}
\end{aligned}
\end{equation}
denote work conjugates of ${{\pmb{\varepsilon }}^{ea}}$. $\sigma _{11}^{e}$, $\sigma _{22}^{e}$, $\sigma _{33}^{e}$, $\sigma _{23}^{e}$, $\sigma _{13}^{e}$, and $\sigma _{12}^{e}$ denote components of the emergent stress tensor, $\Gamma _{23}^{e},\ \Gamma _{13}^{e},\ \Gamma _{12}^{e}$ denote components of the emergent torsion tensor.
Eq. (\ref{8}) describes the linear response of any EC towards small disturbance: for given $d{{\mathbf{X}}^{q}}$ and $d{{\pmb{\sigma }}^{ea}}$, one calculates $d{{\mathbf{v}}^{q}}$ and $d{{\pmb{\varepsilon }}^{ea}}$.
From Eq. (\ref{8}), we also learn that for the considered EC to be a local minimum in the landscape of the free energy functional, the matrix $\pmb{\Phi }=\left[ \begin{matrix}
   {{\mathbf{C}}^{e}} & {{\mathbf{g}}^{eq}}  \\
   {{\left( {{\mathbf{g}}^{eq}} \right)}^{\mathrm T}} & {{\pmb{\mu }}^{q}}  \\
\end{matrix} \right]$ has to be positive-definite. This condition should be guaranteed before any calculation using Eq. (\ref{8}). Moreover, the compliance matrices defined in Eq. (\ref{7}) determines the emergent phonon excitations of ECs at the $\Gamma$ point (i.e., at $\mathbf{k}=\mathbf{0}$), the details of which are explained in a subsequent work of ours\cite{hu2019lagrangian}.

The emergent elastic stresses and emergent torsion introduced in ${{\pmb{\sigma }}^{ea}}$ share the same dimension with the elastic stresses. However, they do not correspond to any kind of macroscopic field that we have known, for which we usually have $d{{\pmb{\sigma }}^{ea}}=\mathbf{0}$. In this case, we have from Eq. (\ref{8})
\begin{equation}
\begin{aligned}
d{{\pmb{\varepsilon }}^{ea}}={{\pmb{\lambda }}}d{{\mathbf{X}}^{q}},
\label{10}
\end{aligned}
\end{equation}
where ${{\pmb{\lambda }}}=-{{\left( {{\mathbf{C}}^{e}} \right)}^{-1}}{{\mathbf{g}}^{eq}}{{\left( {{\pmb{\mu }}^{q*}} \right)}^{-1}}$ and ${{\pmb{\mu }}^{q*}}={{\pmb{\mu }}^{q}}-{{\left( {{\mathbf{g}}^{eq}} \right)}^{\mathrm T}}{{\left( {{\mathbf{C}}^{e}} \right)}^{-1}}{{\mathbf{g}}^{eq}}$. $\pmb{\lambda}$ describes the stiffness of ECs with respect to $\mathbf{X}^q$, the Fourier magnitudes of the primary external field, and is called the ``primary crossover stiffness matrix". The word ``crossover'' means that the matrix links the deformation of an emergent crystalline states with an external field applied to the underlying atomic lattice. $\pmb{\lambda}$ should be distinguished from the emergent elastic stiffness matrix $\mathbf{C}^e$ defined in Eq. (\ref{7}), the latter of which describes the stiffness of ECs when subject to the emergent stress field and emergent torsion field, which do not correspond to any external field we have hitherto known. 
\section{Thermodynamic of emergent crystals in magnetic materials}

Magnetic skyrmions are topologically protected emergent particles. They exist as stable or metastable state in noncentrosymmetric helimagnets due to the competition between the ferromagnetic exchange interaction, favoring a collinear spin alignment, and the DMI, favoring a rotating spin alignment \cite{rossler2006spontaneous}. In experiments, various crystalline states of skyrmions have been observed, including Bloch skyrmions \cite{muhlbauer2009skyrmion,yu2010real,yu2011near}, N\'eel skyrmions \cite{kezsmarki2015neel,kurumaji2017neel,padmanabhan2019optically} and anti-skyrmions \cite{nayak2017magnetic,koshibae2016theory}. They are stabilized by different kinds of DMI permitted in helimagnets with different symmetry. In this section, we first give the free energy density for noncentrosymmetric helimagnets; then by symmetry analysis, we get the mathematical form for different DMI; next, we describe the magnetization structure of skyrmion crystals by Fourier representation; and finally, we solve different magnetization structures of SkX via free energy minimization and study the evolution of SkX with respect to some thermodynamic parameters.

\subsection{Free energy density with DMI for noncentrosymmetric helimagnets}

\begin{table*}
    \caption{Form of $\phi_{\mathrm {DM}}$ for different point groups. $D$, $D'$ and $D''$ are DMI coefficients, $\mathrm {tan} \xi$ describes relative strength of different parts in $\phi_{\mathrm {DM}}$. \cite{bogdanov1989thermodynamically,li2016emergence,gungordu2016stability}}
    \begin{tabular}{c|c|c}
        \hline\hline     
        Point groups & $\phi_{\mathrm {DM}}$ & Types \\
        \hline
        $T,\ O$ & $D(\mathcal{L}_{321}+\mathcal{L}_{132}+\mathcal{L}_{213})$  & Bloch \\
        $D_{3},\ D_{4},\ D_{6}$ &  $D(\mathcal{L}_{321}+\mathcal{L}_{132})+D'\mathcal{L}_{213}$  & Bloch \\
        $C_{3v},\ C_{4v},\ C_{6v}$ & $D(\mathcal{L}_{131}+\mathcal{L}_{232})$  & N\'eel \\
        $D_{2d}$ & $D(\mathcal{L}_{321}-\mathcal{L}_{132})$ &  Anti-I \\
        $C_{3},\ C_{4},\ C_{6}$ & $D[\mathrm {sin}\xi(\mathcal{L}_{321}+\mathcal{L}_{132})+\mathrm {cos}\xi(\mathcal{L}_{131}+\mathcal{L}_{232})]+D'\mathcal{L}_{213}$  &  Bloch-N\'eel mixed \\ 
        $S_4$ & $D[\mathrm {sin}\xi(\mathcal{L}_{321}-\mathcal{L}_{132})+\mathrm {cos}\xi(\mathcal{L}_{131}-\mathcal{L}_{232})]$ & Anti-I-Anti-II mixed \\
        $D_{2}$ & $D(\mathcal{L}_{321}+\mathcal{L}_{132})+D'(\mathcal{L}_{321}-\mathcal{L}_{132})+D''\mathcal{L}_{213}$ &  Bloch or Anti-I \\
        $C_{2v}$ & $D(\mathcal{L}_{131}+\mathcal{L}_{232})+D'(\mathcal{L}_{131}-\mathcal{L}_{232})$ &  N\'eel or Anti-II \\
        \hline\hline
    \end{tabular}
    \label{t1}
\end{table*}

Based on the Landau-Ginzburg mean field theory \cite{bak1980theory}, we write the free energy density for noncentrosymmetric helimagnets in the following form:
\begin{equation}
    \begin{aligned}
        \phi(\mathbf M)=&\sum^3_{i=1}A\left(\frac{\partial \mathbf M}{\partial x_i}\right)^2 +\phi_{\mathrm {DM}}(\mathbf M)-\mathbf B \cdot \mathbf M\\
        &+\phi_{\mathrm L}(\mathbf M).
    \end{aligned}
    \label{e11}
\end{equation} 
Here, the magnetization $\mathbf M$ is chosen as the three-dimensional order parameter field. The first term in Eq. (\ref{e11}) represents the exchange interaction with the stiffness $A$. The second term is the DMI, whose form is closely related to the symmetry of helimagnets. The  third term is the Zeeman coupling to an external magnetic field $\mathbf B$. $\phi _{\mathrm L}(\mathbf M)$ consists of the second and fourth order terms of Landau expansion, it can be expressed as
\begin{equation}
    {{\phi}_{\mathrm{L}}}(\mathbf{M})=\alpha (T-{{T}_{0}}){{\mathbf{M}}^{2}}+\beta {{\mathbf{M}}^{4}},
\end{equation}
where $T_{0}$ is the ordering temperature with zero DMI \cite{leonov2018crossover}, it is related to the ferromagnetic Curie temperature by the formulae $T_{\mathrm{C}}=T_{0}+\frac{D^{2}}{4\alpha A}$ \cite{rossler2006spontaneous,wilhelm2011precursor,wilhelm2012confinement} with $D$ the coefficient reflecting the strength of DMI. \

To get the form of DMI for different helimagnets, we carry out the symmetry analysis. In a continuum model, the free energy density of DMI can be written as a general form  
\begin{equation}
    \phi_{\mathrm {DM}}=\sum_{i,j,k=1}^3 D_{ijk}M_i\frac{\partial M_j}{\partial x_k}
    \label{e1}
\end{equation}
Here, $D_{ijk}$ are the coefficients describing the strength of DMI, $M_i\ (i=1,\ 2,\ 3)$ are components of magnetization vectors, and $x_i\ (i=1,\ 2,\ 3)$ are spatial coordinates. According to the theory of phase transitions by E. M. Lifshitz \cite{landau1980statistical}, DMI can be simplified as a linear summation of Lifshitz invariants 
\begin{equation}
    \mathcal L_{ijk}=M_i\frac{\partial M_j}{\partial x_k}-M_j\frac{\partial M_i}{\partial x_k}.
\end{equation}
As a result, we have $D_{ijk}=-D_{jik}$, and the number of nonzero independent DMI coefficients reduces from 27 to 9. The magnetization $\mathbf M$ is a pseudovector, thus, it transforms under a rotation $\mathbf R$ as $M_{i'}=|\mathbf R| R_{i'i}M_i$. $M_{i'}$ are components of $\mathbf M$ in new Cartesian coordinates; $|\mathbf R|$ is the determinant of $\mathbf R$, for proper (improper) rotation $|\mathbf R|=1\ (-1)$; $R_{i'i}$ is the scalar product of unit vectors along $i$ and $i'$ axes. $\mathbf D$ is a third order tensor; therefore, under a rotation $\mathbf R$, we have 
\begin{equation}
    D_{i'j'k'}=\sum_{i,j,k=1}^3R_{i'i}R_{j'j}R_{k'k}D_{ijk}.
\end{equation}
When $\mathbf R$ is a symmetry operation for the helimagnets, the free energy density of DMI is invariant, thus 
\begin{equation}
    D_{ijk}=D_{i'j'k'}=\sum_{i,j,k=1}^3 R_{i'i}R_{j'j}R_{k'k}D_{ijk}.
    \label{e4}
\end{equation}
By applying the symmetry operations of certain point group to Eq. (\ref{e4}), we can further reduce the number of nonzero independent $D_{ijk}$. In Table \ref{t1}, we list the  free energy density $\phi_{\mathrm {DM}}$ of DMI for different point groups, and we also classify $\phi_{\mathrm {DM}}$ into certain types, including Bloch, N\'eel, Anti-I, Anti-II and two mixed types. About the classification, we will talk about it later.

We use the following rescaling parameters \cite{wan2018exchange}
\begin{equation}
    \begin{aligned}
        &\mathbf r=\frac{\mathbf x} {L_D},\ \mathbf b=\frac{\mathbf B}{B_0},\ \mathbf m=\frac{\mathbf M}{M_0},\  L_D=\frac{2A}{D},\\
        &B_0=2KM_0,\  M_0=\sqrt{\frac{K}{\beta}},\ K=\frac{D^2}{4A}, \\
        &t=\frac{\alpha(T-T_0)}{K},
    \end{aligned}
\end{equation}
to simplify Eq. (\ref{e11}) and get the rescaled free energy density
\begin{equation}
    \begin{aligned}
        \widetilde\phi(\mathbf m)=&\sum^3_{i=1}\left(\frac{\partial \mathbf m}{\partial r_i}\right)^2+\widetilde\phi_{\mathrm {DM}}(\mathbf m)-2\mathbf b \cdot \mathbf m \\
        &+t\mathbf m^2+\mathbf m^4,
    \end{aligned}
    \label{ee14}
\end{equation}
where $\widetilde\phi(\mathbf m)=\frac\beta{K^2}\phi(\mathbf M)$ and $\widetilde\phi_{\mathrm {DM}}(\mathbf m)=\frac\beta{K^2}\phi_{\mathrm {DM}}(\mathbf M)$ are the rescaled total free energy density and the rescaled DMI free energy density, respectively.

\subsection{Fourier representation of magnetization structure}
\begin{table}
    \caption{Information about the Fourier representation for 2D ECs with hexagonal symmetry.}
    \begin{tabular}{c| c c c c c c  }
        
        \hline\hline     
        $i$ & 1 & 2 & 3& 4& 5& 6\\
        \hline
        $n _i$ & 6 & 6 & 6& 12& 6& 6\\  
        $s _i$ & 1 & $\sqrt{3}$ & 2& $\sqrt 7$& 3& $2\sqrt 3$\\
        $\mathbf q_{i1}$ & $[0,1]^{\mathrm T}$ & $[\sqrt 3,0]^{\mathrm T}$ & $[0,2]^{\mathrm T}$ & $[\sqrt 3,2]^{\mathrm T}$& $[0,3]^{\mathrm T}$& $[2\sqrt 3,0]^{\mathrm T}$\\
        \hline\hline

    \end{tabular}
    \label{t2}
\end{table}

In practice, we use the following Fourier representation of 2D ECs instead of Eq. (\ref{2})
\begin{equation}
    \mathbf m=\mathbf m_0+\sum^n_{i=1}\sum^{n_i}_{j=1}\mathbf m_{\mathbf q_{ij}}\mathrm e^{\mathrm i \mathbf q_{ij}\cdot [(\mathbf I- \mathbf F ^e)\mathbf r]}.
    \label{e19}
\end{equation}
When truncated at a specific value of $n$, the $n$th order Fourier representation given in Eq. (\ref{e19}) saves all the significant Fourier terms up to the $n$th order, which is hard to achieve if one uses Eq. (\ref{2}). 

\begin{equation}
    \mathbf F^e=\begin{bmatrix} 
        \varepsilon_{11}^e & \varepsilon_{12}^e+\omega^e \\
        \varepsilon_{12}^e-\omega^e & \varepsilon_{22}^e 
    \end{bmatrix},
\end{equation}
$\varepsilon_{11}^e$ and $\varepsilon_{22}^e$ are the normal strains of SkX,  $\varepsilon_{12}^e$ and $\omega^e$ reflect respectively the shear deformation and rotation of SkX. For 2D ECs distributed in the $xy$ plane, we find that the first six point groups in Table \ref{t1} possess a DMI with higher symmetry, such that a hexagonal or square lattice can be assumed, for which we have $\varepsilon_{11}^e=\varepsilon_{22}^e$ and $\varepsilon_{12}^e$ and $\omega^e$ can be set to be zero. In this case, Eq. (\ref{e19}) reads 
\begin{equation}
    \mathbf m=\mathbf m_0+\sum^n_{i=1}\sum^{n_i}_{j=1}\mathbf m_{\mathbf q_{ij}}\mathrm e^{\mathrm i q \mathbf q_{ij}\cdot \mathbf r},
    \label{ee21}
\end{equation}
with $q=1-\varepsilon_{11}^e=1-\varepsilon_{22}^e$. $\mathbf m_0=[m_{01},\,m_{02},\, m_{02}]^{\mathrm T}$ is the averaged magnetization. $\mathbf q_{ij}\ (j=1,\,2,\,...,\, n_i)$ are the wavecectors of $i$th order waves, there are $n_i$ of them. $\mathbf q_{ij}$ can be seen as vectors of reciprocal lattice spanned by the basis $\mathbf q_{11}$ and $\mathbf q_{12}$, they satisfy the following relations: $|\mathbf q_{i1}|=|\mathbf q_{i2}|=\cdots=|\mathbf q_{in_i}|=s_i$,  $|\mathbf q_{11}|<|\mathbf q_{21}|<\cdots<|\mathbf q_{n1}|$. Without loss of generality, for 2D ECs with hexagonal symmetry,  we set $\mathbf q_{11}=[0,\,1]^{\mathrm T}$ and $\mathbf q_{12}=[-\frac{\sqrt 3}{2},\, -\frac 12]^{\mathrm T}$. Some information about the Fourier representation of hexagonal SkX is listed in Table \ref{t2}. For the description of square SkX, which has also been observed in experiments and in simulations, we set $\mathbf q_{11}=[0,\,1]^{\mathrm T}$ and $\mathbf q_{12}=[-1,\, 0]^{\mathrm T}$. $\mathbf m_{\mathbf q_{ij}}$ denotes the polarization of $\mathbf q_{ij}$ wave. 

\subsection {Decomposition of $\mathbf m_{\mathbf q_{ij}}$}
The free energy functional for 2D ECs can be obtained by substituting Eq.(\ref{ee21}) into Eq. (\ref{ee14}), and performing an integration in space. Here we show that by decomposing $\mathbf m_{\mathbf q_{ij}}$ in an appropriate orthonormal basis, the first six point groups in Table \ref{t1} share exactly the same form of free energy functional for 2D ECs with hexagonal symmetry. The obtained free energy functional is 
\begin{equation}
    \begin{aligned}
        \bar \phi(\mathbf m)=&\frac 1V\int\widetilde\phi(\mathbf m) dV\\
        =&\bar \phi_{\mathrm{per}}+\frac 1V\int\left(\mathbf m^2+\frac{t-1}{2}\right)^2 d V\\
        &+(\mathbf m_0-\mathbf b)^2-\frac{(t-1)^2}{4}-\mathbf b^2,
    \end{aligned}
    \label{e21}
\end{equation}
where $\bar\phi_{\mathrm{per}}=\sum^n_{i=1}\sum^{n_i}_{j=1}(\mathbf m^*_{\mathbf q_{ij}})^{\mathrm T} \mathbf A_{ij}\mathbf m_{\mathbf q_{ij}}$ includes all gradient terms, i.e., the  exchange interaction and DMI which are the dominant parts of the free energy, $(\mathbf m^*_{\mathbf q_{ij}})^{\mathrm T}$ denotes the complex conjugate of $\mathbf m_{\mathbf q_{ij}}$. $\mathbf A_{ij}$ for different point groups have different forms, but they are all Hermitian, and have the same eigenvalues: $\lambda_1=(s_iq-1)^2$, $\lambda_2=(s_iq)^2+1$ and $\lambda_3=(s_iq+1)^2$. In the orthonormal basis spanned by the unit eigenvectors $\mathbf P_{ij1}$, $\mathbf P_{ij2}$ and $\mathbf P_{ij3}$ of $\mathbf A_{ij}$, $\mathbf m_{\mathbf q_{ij}}$ reads
\begin{equation}
    \mathbf m_{\mathbf q_{ij}}=\sum^3_{k=1}c_{ijk}\mathbf P_{ijk},
    \label{ee23}
\end{equation}
where $c_{ijk}=c^{\mathrm {re}}_{ijk}+\mathrm i c^{\mathrm {im}}_{ijk}\,(k=1,\,2,\, 3)$, and $c^{\mathrm {re}}_{ijk}$ and $c^{\mathrm {im}}_{ijk}$ are real variables to be determined. Using Eq. (\ref{ee23}), $\bar\phi_{\mathrm{per}}$ can be written as a simple form 
\begin{equation}
    \bar\phi_{\mathrm{per}}=\sum^n_{i=1}\sum^{n_i}_{j=1}\sum^3_{k=1}\left((c^{\mathrm {re}}_{ijk})^2+(c^{\mathrm {im}}_{ijk})^2\right)\lambda_k.
    \label{e14}
\end{equation}
Obviously, the value of $\bar\phi_{\mathrm{per}}$ is non-negative because $\lambda_k\geq 0$ (the matrix $A_{ij}$ is positive semidefinite). When $q=1$ and $c_{ijk}=0$ for $i\neq 1$ or $k\neq 1$, $\bar\phi_{\mathrm{per}}$ reaches its minimum 0.

We now list the matrix $\mathbf A_{ij}$ and the orthonormal basis for different point groups. For $T$, $O$ or $D_n\, (n=3,\, 4, \, 6)$ point group
\begin{equation}
    \mathbf A_{ij}=\begin{bmatrix}
        1+(s_iq)^2 &0 & 2\mathrm i q_{ijy}\\
        0 &1+(s_iq)^2 & -2 \mathrm i q_{ijx}\\
        -2\mathrm i q_{ijy} & 2\mathrm i q_{ijx} & 1+(s_iq)^2
    \end{bmatrix},
\end{equation}
\begin{equation}
    \begin{aligned}
        \mathbf P_{ij1}=&\frac {1}{\sqrt 2 s_i q}[-\mathrm i q_{ijy},\, \mathrm i q_{ijx},\,s_iq]^{\mathrm T},\\
        \mathbf P_{ij2}=&\frac {1}{s_i q}[q_{ijx},\, q_{ijy},\,0]^{\mathrm T},\\
        \mathbf P_{ij3}=&\frac {1}{\sqrt 2 s_i q}[\mathrm i q_{ijy},\, -\mathrm i q_{ijx},\,s_iq]^{\mathrm T}.
    \end{aligned}
    \label{e23}
\end{equation}
For $C_{nv}\, (n=3,\, 4, \, 6)$ point group
\begin{equation}
    \mathbf A_{ij}=\begin{bmatrix}
        1+(s_iq)^2 &0 & 2\mathrm i q_{ijx}\\
        0 &1+(s_iq)^2 & 2 \mathrm i q_{ijy}\\
        -2\mathrm i q_{ijy} & -2\mathrm i q_{ijx} & 1+(s_iq)^2
    \end{bmatrix},
\end{equation}
\begin{equation}
    \begin{aligned}
        \mathbf P_{ij1}=&\frac {1}{\sqrt 2 s_i q}[-\mathrm i q_{ijx},\, -\mathrm i q_{ijy},\,s_iq]^{\mathrm T} ,\\
        \mathbf P_{ij2}=&\frac {1}{s_i q}[-q_{ijy},\, q_{ijx},\,0]^{\mathrm T},\\
        \mathbf P_{ij3}=&\frac {1}{\sqrt 2 s_i q}[\mathrm i q_{ijx},\, \mathrm i q_{ijy},\,s_iq]^{\mathrm T}. \\
    \end{aligned}
    \label{e27}
\end{equation}
For $D_{2d}$ point group
\begin{equation}
    \mathbf A_{ij}=\begin{bmatrix}
        1+(s_iq)^2 &0 & -2\mathrm i q_{ijy}\\
        0 &1+(s_iq)^2 & -2 \mathrm i q_{ijx}\\
        2\mathrm i q_{ijy} & 2\mathrm i q_{ijx} & 1+(s_iq)^2
    \end{bmatrix}.
\end{equation}
\begin{equation}
    \begin{aligned}
        \mathbf P_{ij1}=&\frac {1}{\sqrt 2 s_i q}[\mathrm i q_{ijy},\, \mathrm i q_{ijx},\,s_iq]^{\mathrm T} ,\\
        \mathbf P_{ij2}=&\frac {1}{s_i q}[-q_{ijx},\, q_{ijy},\,0]^{\mathrm T},\\
        \mathbf P_{ij3}=&\frac {1}{\sqrt 2 s_i q}[-\mathrm i q_{ijy},\, -\mathrm i q_{ijx},\,s_iq]^{\mathrm T}. \\
    \end{aligned}
    \label{e29}
\end{equation}
For $C_{n}\, (n=3,\, 4, \, 6)$  point group
\begin{equation}
    \begin{aligned}
        \mathbf A_{ij}=
        &\begin{bmatrix}
            1+(s_iq)^2 &0 & 2\mathrm i\mathrm {sin}\xi q_{ijy}\\
            0 &1+(s_iq)^2 & -2 \mathrm i \mathrm {sin}\xi q_{ijx}\\
            -2\mathrm i\mathrm {sin}\xi q_{ijy} & 2 \mathrm i \mathrm {sin}\xi q_{ijx} & 1+(s_iq)^2 
        \end{bmatrix}+\\
        &\begin{bmatrix}
            0 &0 & 2\mathrm i\mathrm {cos}\xi q_{ijx}\\
            0 &0 & 2 \mathrm i\mathrm {cos}\xi q_{ijy}\\
            -2\mathrm i\mathrm {cos}\xi q_{ijx} & -2 \mathrm i\mathrm {cos}\xi q_{ijy} & 0 
        \end{bmatrix},\\    
    \end{aligned}
\end{equation}
\begin{equation}
    \begin{aligned}
        \mathbf P_{ij1}=&\frac {1}{\sqrt 2 s_i q}\begin{bmatrix}
            -\mathrm i(\mathrm {sin}\xi q_{ijy}+\mathrm {cos}\xi q_{ijx})\\
            \mathrm i (\mathrm {sin}\xi q_{ijx}-\mathrm {cos}\xi q_{ijy})\\
            s_iq\end{bmatrix},\\
        \mathbf P_{ij2}=&\frac {1}{s_i q}\begin{bmatrix}
            \mathrm {sin}\xi q_{ijx}-\mathrm {cos}\xi q_{ijy}\\
            \mathrm {sin}\xi q_{ijy}+\mathrm {cos}\xi q_{ijx}\\
            0\end{bmatrix},\\
        \mathbf P_{ij3}=&\frac {1}{\sqrt 2 s_i q}\begin{bmatrix}
            \mathrm i(\mathrm {sin}\xi q_{ijy}+\mathrm {cos}\xi q_{ijx})\\
            -\mathrm i (\mathrm {sin}\xi q_{ijx}-\mathrm {cos}\xi q_{ijy})\\
            s_iq\end{bmatrix}.
    \end{aligned}
\end{equation}
For $S_{4}$ point group
\begin{equation}
    \begin{aligned}
        \mathbf A_{ij}=
        &\begin{bmatrix}
            1+(s_iq)^2 &0 & -2\mathrm i\mathrm {sin}\xi q_{ijy}\\
            0 &1+(s_iq)^2 & -2 \mathrm i \mathrm {sin}\xi q_{ijx}\\
            2\mathrm i\mathrm {sin}\xi q_{ijy} & 2 \mathrm i \mathrm {sin}\xi q_{ijx} & 1+(s_iq)^2 
        \end{bmatrix}+\\
        &\begin{bmatrix}
            0 &0 & 2\mathrm i\mathrm {cos}\xi q_{ijx}\\
            0 &0 & -2 \mathrm i\mathrm {cos}\xi q_{ijy}\\
            -2\mathrm i\mathrm {cos}\xi q_{ijx} & 2 \mathrm i\mathrm {cos}\xi q_{ijy} & 0 
        \end{bmatrix},\\    
    \end{aligned}
\end{equation}
\begin{equation}
    \begin{aligned}
        \mathbf P_{ij1}=&\frac {1}{\sqrt 2 s_i q}\begin{bmatrix}
            \mathrm i(\mathrm {sin}\xi q_{ijy}-\mathrm {cos}\xi q_{ijx})\\
            \mathrm i (\mathrm {sin}\xi q_{ijx}+\mathrm {cos}\xi q_{ijy})\\
            s_iq\end{bmatrix},\\
        \mathbf P_{ij2}=&\frac {1}{s_i q}\begin{bmatrix}
            - \mathrm {sin}\xi q_{ijx}-\mathrm {cos}\xi q_{ijy}\\
            \mathrm {sin}\xi q_{ijy}-\mathrm {cos}\xi q_{ijx}\\
            0\end{bmatrix},\\
            \mathbf P_{ij3}=&\frac {1}{\sqrt 2 s_i q}\begin{bmatrix}
            -\mathrm i(\mathrm {sin}\xi q_{ijy}-\mathrm {cos}\xi q_{ijx})\\
            -\mathrm i (\mathrm {sin}\xi q_{ijx}+\mathrm {cos}\xi q_{ijy})\\
            s_iq\end{bmatrix}.
    \end{aligned}
    \label{e33}
\end{equation}
For $D_2$ point group, $\phi_{\mathrm {DM}}= D(\mathcal{L}_{321}+\mathcal{L}_{132})+D'(\mathcal{L}_{321}-\mathcal{L}_{132})$ (for 2D ECs distributed in the $xy$ plane, $\mathcal{L}_{213}=0$). In this case, the ECs that appear in the system are a deformed state of Bloch SkX or Anti-I SkX. If Bloch type is dominant, i.e., $|D|>|D'|$, we chose Eq. (\ref{e23}) as the orthonormal basis, otherwise, we chose Eq. (\ref{e29}). In either case, we have to use Eq. (\ref{e19}) instead of Eq. (\ref{ee21}) to describe the rescaled magnetization. For $C_{2v}$ point group, $\phi_{\mathrm {DM}}=D(\mathcal{L}_{131}+\mathcal{L}_{232})+D'(\mathcal{L}_{131}-\mathcal{L}_{232})$. The ECs are a deformed stete of N\'eel SkX or Anti-II SkX. If $|D|>|D'|$, we chose Eq. (\ref{e27}) as the orthonormal basis, otherwise, we chose Eq. (\ref{e33}) with $\xi=0$. 


\subsection{Diversity of ECs in helimagnets}

\begin{figure*}
    \centering
    \includegraphics[scale=0.72]{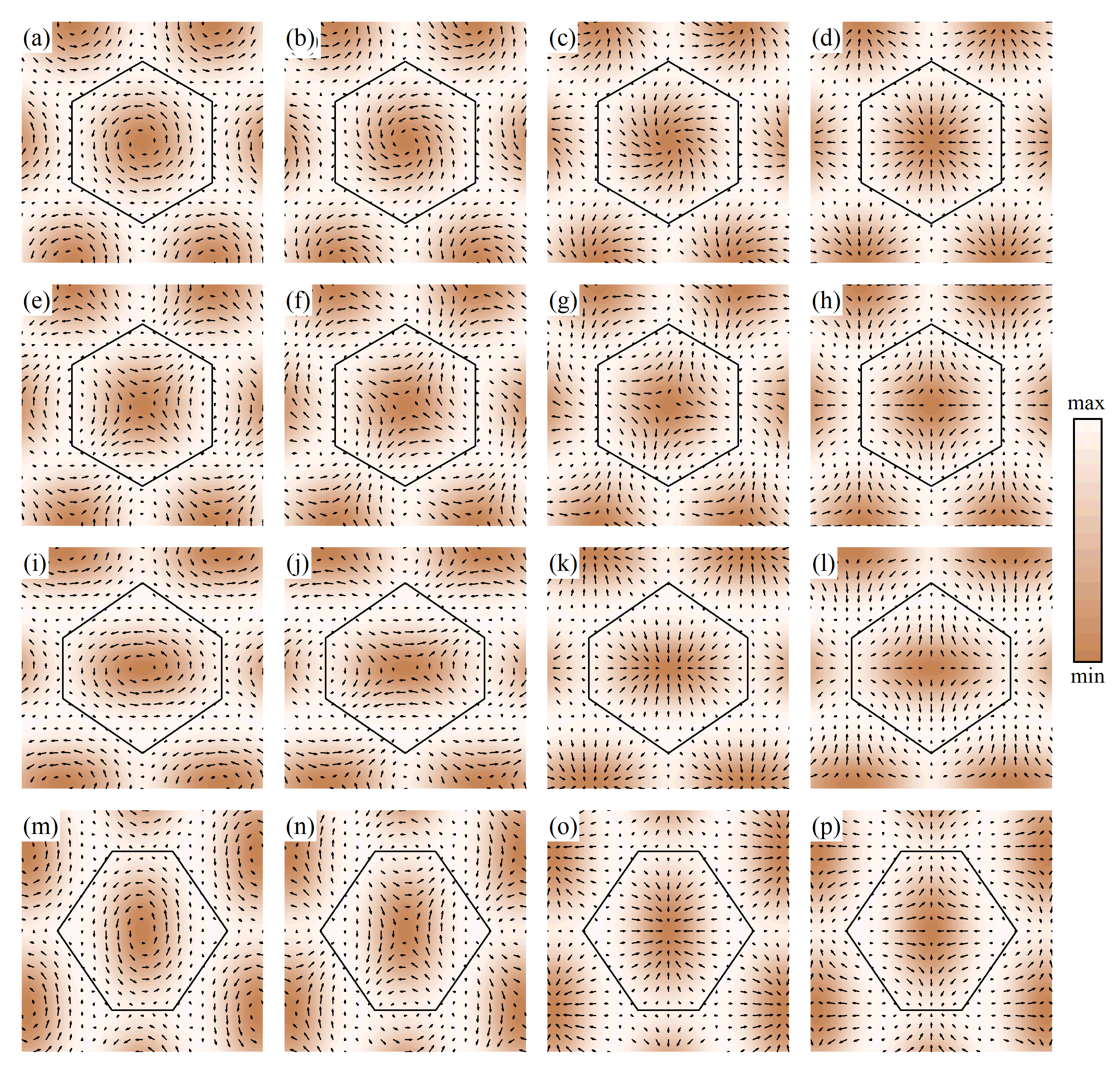}
    \caption{ECs in helimagnets with different point group. (a) Bloch SkX in $T$, $O$ or $D_n \,(n=3,\, 4,\, 6)$ helimagnets. (b-c) Bloch-N\'eel mixed SkX in $C_n \,(n=3,\, 4,\, 6)$ helimagnets with $\xi= 60^\circ$ and $\xi=30^\circ$, respectively. (d) N\'eel SkX in $C_{nv} \,(n=3,\, 4,\, 6)$ helimagnets. (e) Anti-I SkX in $D_{2d}$ helimagnets. (f-g) Anti-I-Anti-II mixed SkX in $S_4$ helimagnets with $\xi= 60^\circ$ and $\xi=30^\circ$, respectively. (h) Anti-II SkX in $S_4$ helimagnets with $\xi= 0^\circ$. (i) and (m) Deformed Bloch SkX in $D_2$ helimagnets with $D'=0.06D$ and $D'=-0.06D$, respectively. (j) and  (n) Deformed Anti-I SkX in $D_2$ helimagnets with $D'=-0.06D$ and $D'=0.06D$, respectively. (k) and (o) Deformed N\'eel SkX in $C_{2v}$ helimagnets with $D'=-0.06D$ and $D'=0.06D$, respectively. (l) and (p) Deformed Anti-II SkX in $C_{2v}$ helimagnets with $D'=-0.06D$ and $D'=0.06D$, respectively. (a-h) are obtained at $t=0$ and $b=0.3$, (i-p) are obtained at $t=0$ and $b=0.4$. The in-plane magnetization components are represented by the arrows, and the out-of-plane magnetization components are illustrated by the colored density plot. The region encircled by the black lines is the Wigner-Seitz cell.}
    \label{f1}
\end{figure*}

The equilibrium states of ECs are determined by free energy minimization at given temperature and magnetic field. FOr all the point groups studied, we find that ECs with hexagonal symmetry always has lower free energy than ECs with square symmetry, for which we focus on the former case. For SkX with hexagonal symmetry, we have
\begin{subequations}
    \begin{align} 
    c_{ijk}&=c_{ilk}  \quad (\mathbf q_{ij}+\mathbf q_{il}\neq 0),\label{8a}\\
    c_{ijk}&=c^*_{ilk}  \quad (\mathbf q_{ij}+\mathbf q_{il}= 0),\label{8b}
    \end{align}
\end{subequations}
while, for deformed SkX, the restrictions Eq. (\ref{8a}) should be discarded. As a result,  $7+\sum^n_{i=1}3n_i$ parameters are needed to describe SkX magnetization texture. They are $\varepsilon_{11}^e,\, \varepsilon_{22}^e,\, \varepsilon_{12}^e,\, \omega^e,\, m_{01},\,m_{02},\,m_{03},\,c^{\text{re}}_{ijk},\,c^{\text{im}}_{ijk} \,(i=1,\,2,\cdots,\,n;\,j=1,\,2,\cdots,\,\frac{n_i}{2};\,k=1,\,2,\,3)$. Hereafter, we focus on the case where $\mathbf b=[0,\, 0,\, b]^{\mathrm T}$. At fixed temperature $t$ and magnetic field $b$, these parameters are obtained by minimizing the rescaled free energy. In this work, the Fourier expansion order is chosen as $n=3$. 



In $C_{nv}\, (n=3,\, 4,\, 6)$ helimagnets, the DMI free energy density $\phi_{\mathrm {DM}}$ can be divided into a Bloch part $\phi_{\mathrm {BL}}=D\mathrm {sin}\xi(\mathcal{L}_{321}+\mathcal{L}_{132})$ and a N\'eel part $\phi_{\mathrm {NE}}=D\mathrm {cos}\xi(\mathcal{L}_{131}+\mathcal{L}_{232})$ (see Table \ref{t1}), where $\xi$ is a parameter characterizing the relative strength of $\phi_{\mathrm {BL}}$ and $\phi_{\mathrm {NE}}$. To investigate the evolution of SkX magnetization structure with respect to $\xi$, we plot Fig. \ref{f1}(a-d) at $\xi=90^\circ,\, 60^\circ,\, 30^\circ,\, 0^\circ$, respectively. At $\xi=90^\circ$ [$\xi=0^\circ$], the in-plane magnetization components are perpendicular [parallel] to the corresponding radial directions. Therefore, the SkX belongs to a Bloch [N\'eel] type, which exists also in helimagnets with $T$, $O$ or $D_n \,(n=3,\, 4,\, 6)$ [$C_{nv} \,(n=3,\, 4,\, 6)$] point group. At $\xi=60^\circ$ or $\xi=30^\circ$, the magnetization structure is between that of Bloch SkX and N\'eel SkX, and we call it Bloch-N\'eel mixed SkX. According to Fig. \ref{f1}(a-d), Bloch SkX transforms into N\'eel SkX by rotating the in-plane magnetization components counterclockwise by $90^\circ$. Similarly, we plot Fig. \ref{f1} (e-h) to illustrate the evolution of SkX in $S_4$ helimagnets from Anti-I type (Fig. \ref{f1}(e)) to Anti-II type (Fig. \ref{f1}(h)). The intermediate states (Fig. \ref{f1}(f) and (g)) are called Anti-I-Anti-II mixed SkX. For Anti-I, which also exists in helimagnets with $D_{2d}$ point group, the in-plane magnetization components along a $\langle 0\, 1\rangle$ [$\langle 1\, 1\rangle$] axis are perpendicular [parallel] to the radial direction. For Anti-II SkX,  the opposite is the case. The in-plane magnetization components along a $\langle 0\, 1\rangle$  [$\langle 1\, 1\rangle$] axis are parallel [perpendicular] to the radial direction. The Anti-I to Anti-I transformation is also accomplished by rotating the in-plane magnetization components counterclockwise by $90^\circ$. 

It has been proved by numerical simulation that anisotropic DMI deforms isolated skyrmion from a circular one to an elliptic one \cite{huang2017stabilization}. Here, we show that anisotropic DMI, which is present in $D_2$ and $C_{2v}$ helimagnets, can also deform SkX. We first set $D'=0.06D$ and $D'=0.06D$  for $\phi_{\mathrm{DM}}$ of $D_2$ point group (see Table \ref{t1}) and plot the magnetization distribution of SkX in Fig. \ref{f1}(i) and (m). In this case, SkX belongs to a Bloch type. Due to the existence of Anti-I type DMI, the shape of a skyrmion cell is no longer a regular hexagon and the core of the skyrmion cell is elliptic. Then we consider the other case where Anti-I type DMI dominants and set $D=-0.06D'$ and $D=0.06D'$. The SkX is now a deformed Anti-I type (Fig. \ref{f1}(j) and (n)). Similarly, deformed N\'eel type SkX and deformed Anti-II type SkX in $D_2$ helimagnets are plotted in Fig. \ref{f1} (k) (o) (l) and (p) for $D'=-0.06D$, $D'=0.06D$, $D=-0.06D'$ and $D=0.06D$, respectively.

When we plot Fig. \ref{f1}(a-h) (and (i-p)), two phenomena attract our attention. The first one is that the skyrmion cells have the same size, the second one is that the out-of-plane magnetization components have the same maximum and minimum. To explain these phenomena, we first compare the analytical expressions of the free energy with different types of DMI. It is found that if $m_{01}=m_{02}=0$, i.e., the magnetic field is applied along the $z$ axis, the analytical expressions of free energy are the same. This means that when we do not consider in-plane anisotropy induced by tilted magnetic field, different kinds of SkX can be studied in a unified framework. By minimizing the free energy expressed in Eq. (\ref{e21}) at certain temperature and magnetic field, we can obtain the same set of values for the parameters $\varepsilon_{11}^e,\, \varepsilon_{22}^e,\, \varepsilon_{12}^e,\, \omega,\, m_{01},\,m_{02},\,m_{03},\,c^{\text{re}}_{ijk},\,c^{\text{im}}_{ijk} \,(i=1,\,2,\cdots,\,n;\,j=1,\,2,\cdots,\,\frac{n_i}{2};\,k=1,\,2,\,3)$ for different types of DMI, including Bloch, N\'eel, Anti-I, Anti-II, Bloch-N\'eel mixed and Anti-I-Anti-II mixed. As to the second phenomenon, we express analytically the out-of-plane magnetization components for different types of DMI and find that they are the same. Therefore, different kinds of SkX have the same distribution of out-of-plane magnetization. That is the reason why the second phenomenon
occurs.  

In Ref. \cite{rowland2016skyrmions}, it is shown that when uniaxial anisotropy, which has the form $m_z^2$, is present, 
N\'eel SkX has a larger stable region in the phase diagram than Bloch SkX. However, according to our results, with uniaxial anisotropy considered, the free energy is still the same for Bloch SkX and N\'eel SkX, because the out-of-plane magnetization distribution is the same for these two kinds of SkX. Uniaxial anisotropy does not favor Bloch SkX or N\'eel SkX, but it enlarges the stable region of N\'eel SkX by suppressing the conical phase. 

\subsection{Evolution of SkX with respect to some thermodynamic parameters}
\begin{figure}
    \centering
    \includegraphics[scale=0.55]{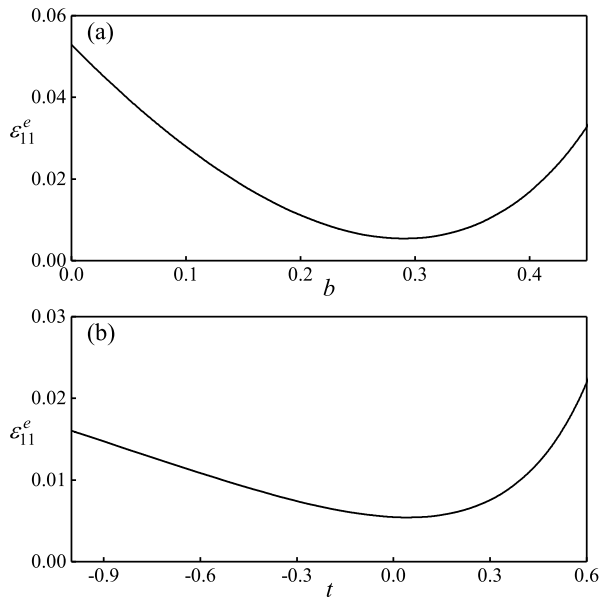}
    \caption{(a)$\varepsilon_{11}^e$ as a function of the magnetic field $b$ at the temperature $t=0$. (b) $\varepsilon_{11}^e$ as  function of the temperature $b$ at the magnetic field $b=0.3$.}
    \label{f2}
\end{figure}

Deformation of SkX consists of two aspects, the shape deformation reflected by the parameters $\varepsilon_{11}^e,\, \varepsilon_{22}^e,\, \varepsilon_{12}^e,\, \omega$ (called lattice deforamtion), and the deformation reflected by the inequality of $c_{ijk}\, (j=1,\, 2,\, \cdots,\, \frac{n_i}{2})$ (called in-lattice deformation). When there is no in-plane anisotropy, such as DMI anisotropy and anisotropy induced by tilted magnetic field, SkX has hexagonal symmetry. In this case, no in-lattice deformation occurs, because the waves along $\mathbf q_{ij}\, (j=1,\, 2,\, \cdots,\, \frac{n_i}{2})$ directions are equivalent, and $c_{ijk}\, (j=1,\, 2,\, \cdots,\, \frac{n_i}{2})$ reflecting the wave amplitude are equal. About the lattice deformation, we have $\varepsilon_{11}^e=\varepsilon_{22}^e$, $\varepsilon_{12}^e=0$, and $\omega^e=0$, for which the only parameter is $\varepsilon_{11}^e$.

Consider Bloch type DMI without anisotropy and apply a magnetic field perpendicular to the skyrmion plane, we study the evolution of the normal strain $\varepsilon_{11}^e$  with respect to the thermodynamic variables $b$ and $t$. We first fix the temperature $t=0$, and study the influence of magnetic field $b$ on the size of a skyrmion cell, the result is plotted in Fig. \ref{f2}(a). It is found that with increasing $b$,  $\varepsilon_{11}^e$ decreases from 0.053 (at $b=0$)  to it's minimum 0.0054 (at $b=0.29$) then increase to 0.042 (at $b=0.45$). We then fix the magnetic field $b=0.3$, and study the thermal expansion of SkX. As shown in Fig. \ref{f2}(b), $\varepsilon_{11}^e$ decreases for $-1<t<0.04$ then increases for $0.04<t<0.6$ with increasing temperature. This means that the coefficient of thermal expansion is negative for $-1<t<0.04$ and it changes to be positive for $0.04<t<0.6$.

\begin{figure}
    \centering
    \includegraphics[scale=0.55]{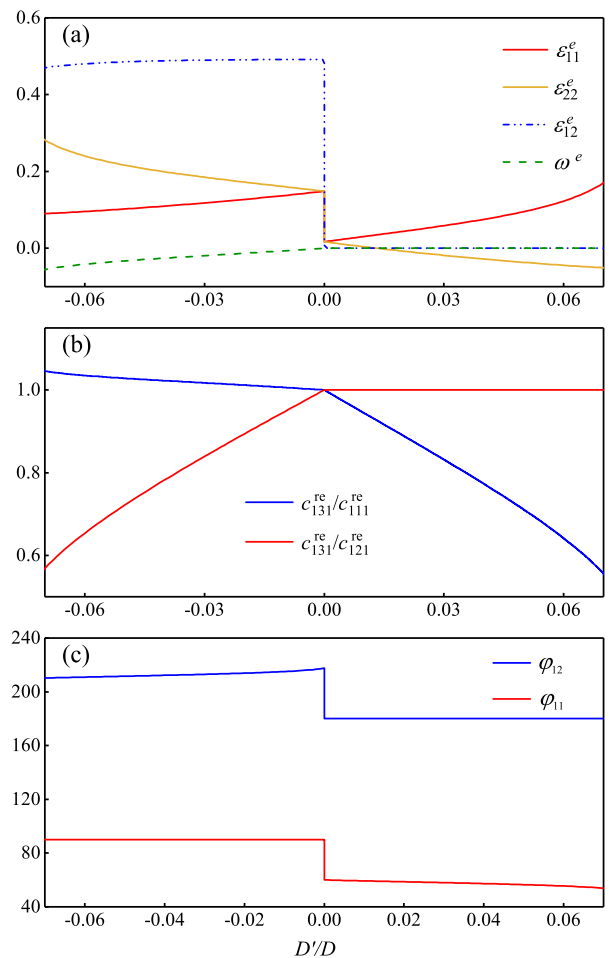}
    \caption{(a) $\varepsilon_{11}^e$ and $\varepsilon_{22}^e$, (b) $c_{131}^{\mathrm {re}}/c_{111}^{\mathrm {re}}$ and $c_{131}^{\mathrm {re}}/c_{121}^{\mathrm {re}}$, and (c) $\varphi_{11}$ and $\varphi_{12}$ as functions of $D'/D$. The results are calculated at the magnetic field $b=0.4$ and the temperature $t=0$.}
    \label{f3}
\end{figure}

We now study the influence of DMI anisotropy on the deformation of SkX in $D_2$ or $C_{2v}$ helimagnets. The DMI free energy density considered is $\phi_{DM}=D(\mathcal{L}_{321}+\mathcal{L}_{132})+D'(\mathcal{L}_{321}-\mathcal{L}_{132})$ with Bloch type DMI the dominant part, i.e., $|D'/D|<1$. We fix the temperature $t=0$ and the magnetic field $b=0.4$. The lattice-deformation-related parameters $\varepsilon_{11}^e$, $\varepsilon_{22}^e$, $\varepsilon_{12}^e$ and $\omega$ as functions of $D'/D$, which characterize the strength of DMI anisotropy, are plotted in Fig. \ref{f3}(a). The curves of  $\varepsilon_{11}^e$, $\varepsilon_{22}^e$ and $\omega$ are discontinuous at $D'/D=0$. This means that a phase transition happens when $D'/D$ changes its sign. To work out what happens during this phase transition, we plot Fig. \ref{f3} (c), which shows $\varphi_{11}$ and $\varphi_{12}$, the angles between the deformed wave vectors $\mathbf q_{11}^d$, $\mathbf q_{12}^d$ and the $x$ axis, as functions of $D'/D$. We can see that $\varphi_{11}$ and $\varphi_{12}$ jump at $D'/D=0$, i.e., a rotation of SkX occurs. Actually, the magnetization structure for a negative $D'/D$ can be seen as a $90^\circ$ rotation of the magnetization structure for a corresponding positive $D'/D$. Because, the DMI anisotropy free energy changes its sign under $90^\circ$ rotation.


As to the in-lattice deformation, it is mainly reflected by the first order wave amplitudes $c_{111}^{\mathrm {re}}$, $c_{121}^{\mathrm {re}}$ and $c_{131}^{\mathrm {re}}$. $c_{121}^{\mathrm {re}}/c_{111}^{\mathrm {re}}$ and $c_{121}^{\mathrm {re}}/c_{131}^{\mathrm {re}}$ as functions of $D'/D$ are plotted in Fig. \ref{f3}(b). For positive $D'/D$, the value of $c_{121}^{re}/c_{131}^{re}$ are always 1, meaning that the waves along deformed $\mathbf q_{11}$ and $\mathbf q_{12}$ directions are equivalent. This is reasonable, because the DMI anisotropy do not break the antisymmetry of DMI with respect to the $y$ axis.

\section{Elasticity of emergent crystals in magnetic materials under bias magnetic fields}

Now we study the emergent elasticity for all the ECs that appear in helimagnets when subject to a disturbance of the bias magnetic field. In other words, we try to derive the linear relationship between $d\pmb \varepsilon^{ea}$ and $d\mathbf{b}=[db_1, db_2, db_3]^{\mathrm T}$. Compared with the generalized relation given in Eq. (\ref{10}), we further assume that spatially periodic magnetic fields are not applied (i.e., the work conjugates of  $c_{ijk}$ are zero). In this case,  Eq. (\ref{10}) changes to

\begin{equation}
    \begin{bmatrix} 
        d\varepsilon_{11}^e \\ d\varepsilon_{22}^e \\d\varepsilon_{12}^e \\d\omega^e \end
    {bmatrix} 
    =
    \pmb \lambda^{bias}
    \begin{bmatrix}
        db_1 \\ db_2\\ db_3
    \end{bmatrix},
\end{equation}
where $\pmb \lambda^{bias}$ is a $4*3$ matrix whose components depend on the temperature $t$ and magnetic field $b$, $db_i\, (i=1,\, 2,\, 3)$ are small distributions of the bias magnetic field. $\lambda_{ij}$ can be expressed analytically, but their expressions are too length to be present. Here, we just calculate their numerical values. 

\begin{figure}
    \centering
    \includegraphics[scale=0.32]{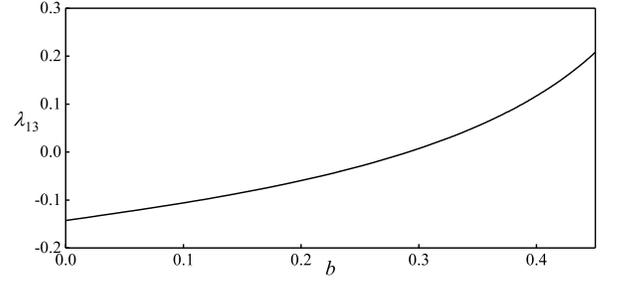}
    \caption{$\lambda_{13}$ as a function of $b$ at $t=0$ for ECs permitted by the first six types of point groups in Table \ref{t1}.}
    \label{f4}
\end{figure}

Based on symmetry analysis, we find that for the ECs permitted by the first six types of point groups listed in Table \ref{t1}, we have

\begin{equation}
    \pmb \lambda^{bias}=
    \begin{bmatrix}
        0 & 0 & \lambda_{13} \\
        0 & 0 & \lambda_{13} \\
        0 & 0 & 0 \\
        0 & 0 & 0 \\
    \end{bmatrix},
\end{equation}
which means that without in-plane anisotropy, the bias magnetic field can only induce normal strain of the ECs, and small disturbance of in-plane bias magnetic field does not induce lattice deformation. $\lambda_{13}$ as a function of $b$ is plotted in Fig. \ref{f4}. It is shown that with increasing magnetic field, $\lambda_{13}$ increases from a negative value to a positive value, and at $b=0.29$, $\lambda_{13}=0$. This accords with the results shown in Fig. \ref{2}(a). Actually, $\lambda_{13}$ represents the slope of the $\varepsilon_{11}^e-b$ curve. 

\begin{figure}
    \centering
    \includegraphics[scale=0.56]{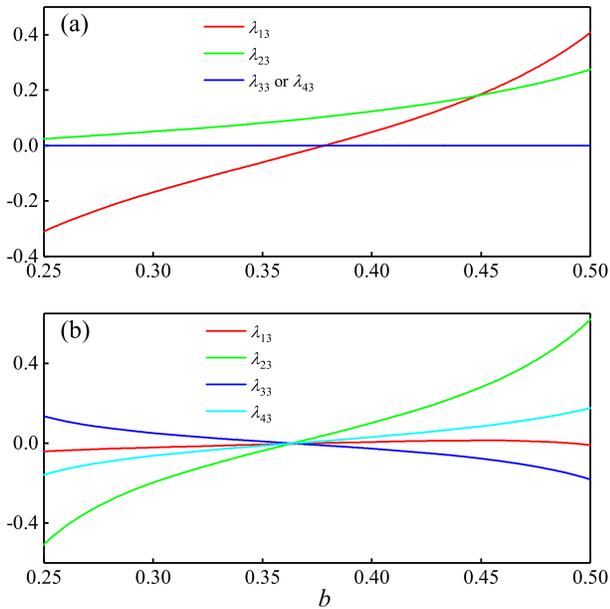}
    \caption{$\lambda_{13}$, $\lambda_{23}$, $\lambda_{33}$ and $\lambda_{43}$ as functions of $b$ at $t=0$ for $D_2$ point group. (a) $D'/D=0.03$  and  (b) $D'/D=-0.03$.}
    \label{f5}
\end{figure}

Meanwhile, we find that for the ECs permitted by the last two types of point groups listed in Table \ref{t1}, we have
\begin{equation}
    \pmb \lambda^{bias}=
    \begin{bmatrix}
        0 & 0 & \lambda_{13} \\
        0 & 0 & \lambda_{23} \\
        0 & 0 &\lambda_{33} \\
        0 & 0 &\lambda_{43} \\
    \end{bmatrix}.
\end{equation}
In this case,  $\lambda_{13}$ is no longer equal to $\lambda_{23}$, meaning that anisotropic lattice deformation of ECs takes place. Fig. \ref{f5} shows $\lambda_{i3}\, (i=1,\, 2,\, 3,\, 4)$ of $D_2$ helimagnets as functions of the magnetic field $b$ at the temperature $t=0$. When $D'/D$ is positive (Fig. \ref{f5} (a)), $\lambda_{33}$ and $\lambda_{43}$ are zero. Therefore, the bias magnetic field does not change the value of $\varepsilon_{12}$ and $\omega$. When $D'/D$ is negative (Fig. \ref{f5} (b)),   $\lambda_{i3}\, (i=1,\, 2,\, 3,\, 4)$ all vary with respect to the magnetic field. We should emphasis that for $D_2$ and $C_{2v}$ helimagnets, if the orthonormal basis and the sign of $D'/D$ or $D/D$ are appropriately chosen, the free energy functional share the same form in terms of $c_{ijk}$ and $\pmb \varepsilon^{ea}$; therefore, $\pmb \lambda^{bias}$ of $D_2$ and $C_{2v}$ helimagnets behave similarly with respect to the thermodynamic parameters.

\begin{acknowledgments}
    The work was supported by the NSFC (National Natural Science Foundation of China) through the funds 11772360, 11472313, 11572355 and Pearl River Nova Program of Guangzhou (Grant No. 201806010134).
\end{acknowledgments}

\bibliographystyle{apsrev4-1}
\bibliography{Manuscript}

\end{document}